\documentclass[aps,prl,twocolumn,showpacs,preprintnumbers,amsmath,amssymb,floatfix]{revtex4-2}

\usepackage{graphicx}
\usepackage{longtable}
\usepackage{dcolumn}
\usepackage{bm}
\usepackage{color}
\usepackage[normalem]{ulem}
\usepackage[colorlinks=true, urlcolor=blue, linkcolor=blue, citecolor=blue]{hyperref}
\usepackage[all]{hypcap}
\newcommand{\beq}{\begin{equation}}
\newcommand{\eeq}{\end{equation}}
\newcommand{\bea}{\begin{eqnarray}}
\newcommand{\eea}{\end{eqnarray}}

\begin{document}


\title{Chiral-Angle-Controlled Spin Splitting and Spin Transport in Nanotubes Rolled from d-wave Altermagnets}

\author{Ersoy \c{S}a\c{s}{\i}o\u{g}lu$^{1}$}\email{ersoy.sasioglu@physik.uni-halle.de}
\author{Tom G. Saunderson$^{1}$}
\author{B\"{o}rge G\"{o}bel$^{1}$}
\author{Ingrid Mertig$^{1}$}
\author{Samir Lounis$^{1}$}

\affiliation{$^{1}$Institute of Physics and Halle-Berlin-Regensburg Cluster of Excellence CCE, Martin Luther University Halle-Wittenberg, 06120 Halle (Saale), Germany}


\begin{abstract}

Altermagnets combine compensated collinear magnetic order with
momentum-dependent spin splitting in the electronic band structure.
Here, we show that rolling a two-dimensional (2D) $d$-wave altermagnet
into a nanotube converts this momentum dependence into
chiral-angle-controlled one-dimensional (1D) spin splitting through
dimensional projection. A minimal tight-binding model reveals a
characteristic nodal--antinodal dependence on the chiral angle $\theta$,
with the central circumferential subband exhibiting a $\cos(2\theta)$
scaling and the projected spin splitting vanishing for the nodal
orientation and reversing sign between orthogonal antinodal
orientations. First-principles calculations for V$_2$O and
representative symmetric and Janus systems demonstrate that this
nodal--antinodal selection rule persists despite curvature-induced
structural asymmetry and magnetic moment imbalance. We further show
that the projected electronic structure produces
chiral-angle-controlled spin-polarized transport: antinodal nanotubes
exhibit spin-polarized transmission, whereas the nodal nanotube remains
conducting with identical spin-channel transmission. These results
demonstrate how dimensional projection can translate the momentum-space
spin splitting of a 2D altermagnet into geometrically controlled
electronic and transport properties in nanotubes.

\end{abstract}

\maketitle

Altermagnets have recently emerged as a distinct class of collinear
magnetic materials that exhibit momentum-dependent spin splitting despite
compensated magnetic order
~\cite{Zunger,Smejkal2020_SciAdv,Smejkal2022_PRX,Smejkal2022,
Smejkal2022Nature,Jungwirth2024}. Unlike conventional ferromagnets, where
exchange-induced spin splitting is approximately uniform in momentum
space, altermagnetic spin splitting changes sign between symmetry-related
regions of the Brillouin zone and can form highly anisotropic $d$-, $g$-,
or higher-order wave patterns
~\cite{Mazin2023_MnTe,Giuli2024_NodalReview}. The coexistence of sizable
spin splitting with compensated magnetic order has stimulated
considerable interest in altermagnets as platforms for spin-dependent
electronic phenomena with strongly reduced macroscopic stray fields
~\cite{GonzalezHernandez2021,Feng2022,Jungwirth2025_AltermagSpintronics}.
Momentum-dependent spin splitting has now been observed experimentally
in several candidate materials, including MnTe, CrSb, and
KV$_2$Se$_2$O, through angle-resolved photoemission spectroscopy and
complementary probes
~\cite{Lee2024_PRL_MnTe,Reimers2024_NatCommun_CrSb,
Jiang2025_NatPhys_KV2Se2O}. These developments have been accompanied by
rapid progress in symmetry classification, materials discovery, and
first-principles studies of bulk and low-dimensional altermagnets,
including an expanding family of two-dimensional (2D) systems
~\cite{yuan2020giant,sodequist2024two,jungwirth2026symmetry,
bhattarai2025high,bai2024altermagnetism,Tamang_Review_Altermagnets,
mcclarty2024landau,song2025electrical,Brahimi2024_RuO2_film,
fukaya2025superconducting}.

A defining feature of many altermagnetic states is the presence of
symmetry-protected momentum-space directions along which the spin
splitting vanishes. In square-lattice 2D $d$-wave altermagnets, this
structure takes a particularly transparent form: a
$d_{x^2-y^2}$-type spin splitting produces antinodal directions with
large and opposite-sign splitting, separated by nodal directions where
the bands remain spin degenerate. This nodal--antinodal structure raises
a more general question concerning dimensional reduction. When a
higher-dimensional electronic structure is confined to one dimension,
transverse momentum becomes quantized while dispersion remains along the
one-dimensional propagation direction. Different geometrical
orientations can therefore sample distinct sectors of the original
momentum-dependent electronic structure. This suggests that dimensional
projection could select between nodal and antinodal sectors of a
$d$-wave altermagnet and thereby convert momentum-space spin anisotropy
into a geometrically controlled one-dimensional spin splitting.

\begin{figure*}[!ht]
\centering
\includegraphics[width=0.85\textwidth]{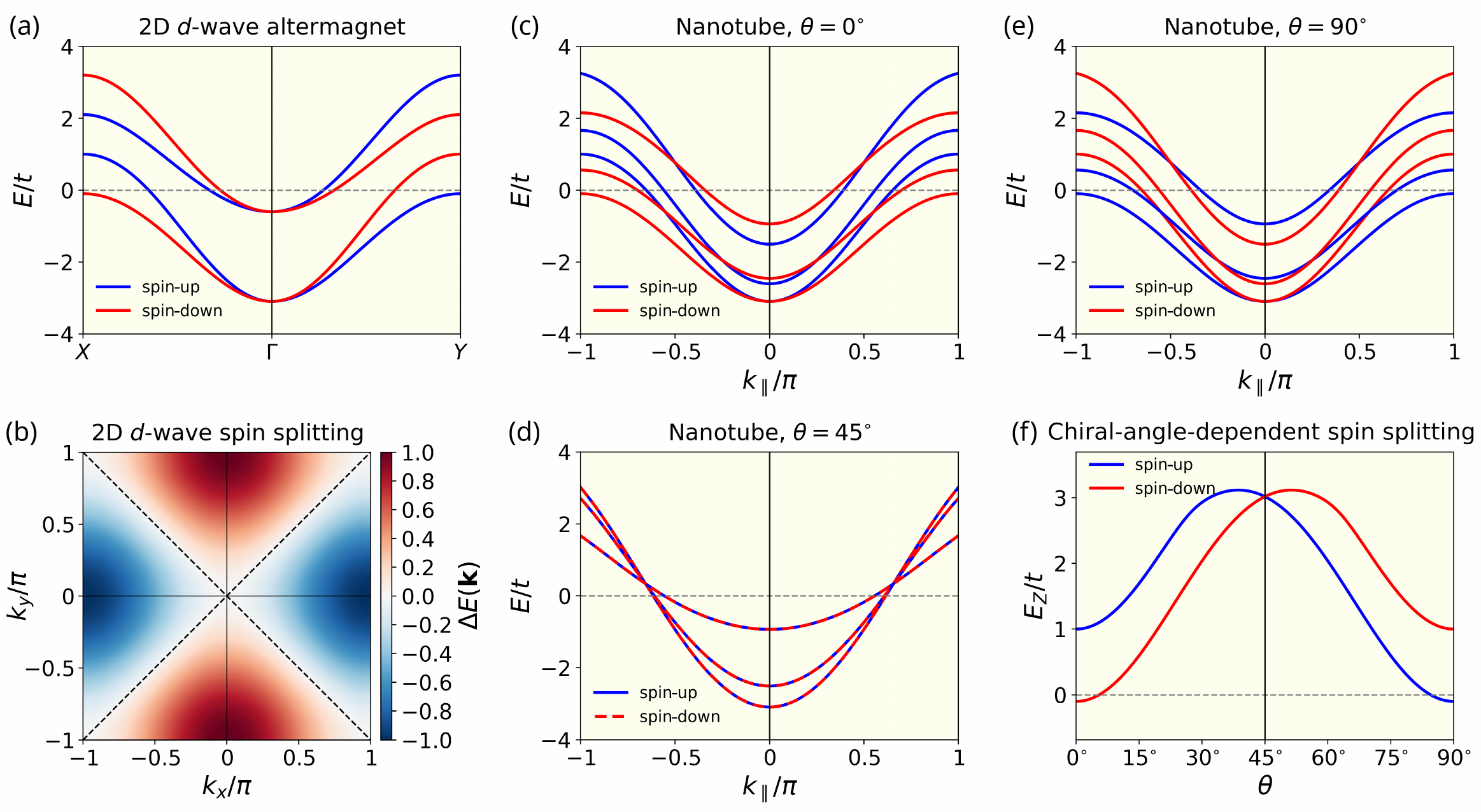}
\vspace{-0.4 cm}
\caption{\textbf{Dimensional projection of $d$-wave spin splitting into one-dimensional nanotube states}.
(a) Band structure of the parent 2D $d$-wave altermagnet along
X--$\Gamma$--Y.
(b) Momentum-resolved spin splitting $\Delta E(\mathbf{k})$ of the
parent system, showing antinodal regions with opposite signs of the spin
splitting separated by symmetry-protected nodal lines.
(c)--(e) Spin-resolved 1D band structures for nanotubes with chiral
angles $\theta=0^\circ$, $45^\circ$, and $90^\circ$, respectively.
These three limiting geometries are achiral. Antinodal orientations
produce finite spin splitting with opposite signs for $\theta=0^\circ$
and $90^\circ$, whereas the nodal orientation at $\theta=45^\circ$
yields spin-degenerate bands.
(f) Spin-resolved energies of the $k_{\perp}=0$ subband at the nanotube
zone-boundary point $Z$ as a function of chiral angle $\theta$.
The
splitting evolves continuously between the two antinodal orientations,
vanishes at the nodal orientation, and reverses sign, following the
$\cos(2\theta)$ scaling of the projected $d$-wave form factor for this
circumferential mode.}
\label{fig1}
\end{figure*}

Nanotubes provide a natural setting in which to realize this principle.
Rolling a 2D layer into a cylinder imposes periodic boundary conditions
around the circumference, quantizing the transverse momentum and
producing a set of one-dimensional (1D) subbands dispersing along the
nanotube axis. In carbon nanotubes, the wrapping geometry determines the
allowed momentum-space cuts through the graphene band structure and
thereby whether the resulting nanotube is metallic or semiconducting
~\cite{Saito1998,Dresselhaus2001}. Related dimensional-confinement
effects arise more broadly in nanotubes formed from layered materials,
for which substantial experimental progress has been achieved in
transition-metal dichalcogenides and Janus systems
~\cite{Nakanishi2023,Liu2021,Dai2025,Yang2025,mikkelsen2021band,
zhao2023curvature}. Recent theoretical work has also explored
altermagnetic nanotubes in the context of Lifshitz-transition-induced
open Fermi contours and chirality-dependent persistent spin currents
~\cite{Kochan2026}. Together with the growing set of predicted 2D
$d$-wave altermagnets~\cite{Zou2024,sodequist2024two,Li2024,Biswas2026}, these
developments establish altermagnetic nanotubes as a promising setting
for geometrically controlled spin-dependent phenomena. Here, we focus
instead on how the sign-changing momentum-dependent spin splitting of
the parent $d$-wave state is transferred to the one-dimensional
electronic structure through dimensional projection. The nanotube
geometry can thereby select different sectors of the nodal--antinodal
structure, transforming momentum-space magnetic anisotropy into a
real-space geometrical control parameter.

Whether this projected spin splitting also controls transport is a
separate question. A spin-split band structure does not by itself imply
a spin-polarized current, because transport depends on the propagating
states available near the Fermi level. Establishing whether the
nodal--antinodal structure survives at the level of the conducting
channels is therefore essential for connecting dimensional projection
to spintronic functionality. Nanotubes are particularly appealing in
this respect because their geometry is structurally fixed and can, in
principle, control the relative contributions of the two spin channels
without changing the underlying magnetic order. Whether such
geometrical control can be realized in realistic nanotubes rolled from
2D $d$-wave altermagnets remains unexplored.

Here, we establish dimensional projection as a mechanism for converting
the momentum-dependent spin splitting of a 2D $d$-wave altermagnet into
geometrically controlled 1D electronic and transport properties. A
minimal tight-binding model reveals a characteristic nodal--antinodal
dependence on the chiral angle $\theta$, with the central circumferential
subband exhibiting a $\cos(2\theta)$ scaling of the projected spin
splitting. We then verify the resulting nodal--antinodal selection rule
using first-principles calculations for V$_2$O and structurally distinct
symmetric and Janus nanotubes, showing that it persists despite
curvature-induced structural asymmetry and magnetic moment imbalance.
Finally, we demonstrate that this geometrical dependence is reflected
in spin-polarized transport, with the spin-channel imbalance
progressively suppressed toward the nodal orientation while finite
charge transmission is retained.

\section*{Results}

\subsection*{Dimensional projection of $d$-wave spin splitting}

We first establish the general mechanism by which the momentum-dependent
spin splitting of a 2D $d$-wave altermagnet is converted into
angle-dependent one-dimensional electronic states upon rolling. To
separate the underlying geometrical and symmetry principles from
material-specific effects, we consider a minimal two-sublattice
tight-binding model on a square lattice describing a generic metallic
$d$-wave altermagnet. The model provides an analytical framework for
determining how the orientation of the nanotube axis relative to the
nodal--antinodal structure of the parent system controls the projected
1D spin splitting. Model parameters and further numerical details are
provided in the Supplementary Information.

The Bloch Hamiltonian is
\begin{equation}
H_{\sigma}(\mathbf{k})
=
\varepsilon(\mathbf{k})\tau_{0}
+t_{AB}(\mathbf{k})\tau_{x}
+
\left[
J\sigma
+
\lambda g_{d}(\mathbf{k})
\right]\tau_{z},
\label{eq:Hk}
\end{equation}
where $\sigma=\pm1$ denotes the spin index, $\tau_i$ are Pauli matrices
acting in the magnetic sublattice space,
$\varepsilon(\mathbf{k})$ describes the spin-independent dispersion,
$t_{AB}(\mathbf{k})$ is the intersublattice hopping, $J$ is the
staggered exchange field, and $\lambda$ controls the strength of the
$d$-wave term. The characteristic momentum dependence of the $d$-wave
state is encoded in the form factor
\begin{equation}
g_{d}(\mathbf{k})
=
\cos k_x-\cos k_y,
\label{eq:dwave}
\end{equation}
which changes sign under a $\pi/2$ rotation and vanishes along
$k_x=\pm k_y$. The resulting spin splitting therefore has two
orthogonal antinodal directions with opposite signs, separated by nodal
lines along which the bands remain spin degenerate. This
$d_{x^2-y^2}$ structure is illustrated by the parent 2D band structure
and momentum-resolved spin splitting in
Figs.~\ref{fig1}(a) and \ref{fig1}(b), respectively.

Rolling the 2D system into a nanotube imposes periodic boundary
conditions around the circumference and restricts the parent electronic
structure to a discrete set of one-dimensional modes. We define the
chiral angle $\theta$ by the orientation of the nanotube axis relative
to an antinodal direction of the parent layer and introduce momentum
components $(k_{\parallel},k_{\perp})$ parallel and perpendicular to
the nanotube axis, respectively. The circumferential component
$k_{\perp}$ is quantized, whereas $k_{\parallel}$ remains the continuous
crystal momentum along the nanotube axis, following the standard
nanotube zone-folding construction
~\cite{Saito1998,Dresselhaus2001}. Changing $\theta$ therefore changes
which sectors of the parent nodal--antinodal structure contribute to
the 1D dispersion.

This geometrical dependence follows directly from the transformation of
the $d$-wave form factor. Expanding Eq.~(\ref{eq:dwave}) around the
$\Gamma$ point and expressing the momentum in the rotated nanotube
coordinates gives
\begin{equation}
g_d
\propto
\cos(2\theta)
(k_{\parallel}^2-k_{\perp}^2)
-
2\sin(2\theta)
k_{\parallel}k_{\perp}.
\label{eq:rotated}
\end{equation}
For the $k_{\perp}=0$ circumferential mode, the projected spin splitting
therefore scales as
\begin{equation}
\Delta_{\rm tube}(\theta)
\propto
\cos(2\theta)\,k_{\parallel}^2.
\label{eq:cos2theta}
\end{equation}
Equation~(\ref{eq:cos2theta}) provides a simple analytical description
of the nodal--antinodal angular dependence for this projected subband:
the spin splitting is maximal in magnitude for the antinodal
orientations $\theta=0^\circ$ and $90^\circ$, vanishes for the nodal
orientation $\theta=45^\circ$, and reverses sign between the two
orthogonal antinodal directions. For the remaining quantized
circumferential modes with $k_{\perp}\neq0$, additional terms contribute,
so that the spin splitting does not in general follow a universal
$\cos(2\theta)$ dependence.

The corresponding nanotube band structures are shown in
Figs.~\ref{fig1}(c)--\ref{fig1}(e). The three representative angles
$\theta=0^\circ$, $45^\circ$, and $90^\circ$ correspond to achiral
nanotube geometries, while intermediate values of $\theta$ generally
describe chiral nanotubes. For $\theta=0^\circ$, projection along an
antinodal direction produces pronounced spin splitting of the 1D bands
[Fig.~\ref{fig1}(c)]. At $\theta=45^\circ$, the nanotube axis is aligned
with a nodal direction and the bands become spin degenerate
[Fig.~\ref{fig1}(d)]. Rotating further to the orthogonal antinodal
orientation at $\theta=90^\circ$ restores finite spin splitting with
the opposite sign [Fig.~\ref{fig1}(e)]. These three achiral orientations
thus sample the positive, zero, and negative sectors of the parent
$d$-wave spin-splitting pattern.

Dimensional reduction also folds states from different momenta of the
parent 2D Brillouin zone into the reduced 1D Brillouin zone of the
nanotube. Consequently, spin-split states originating at finite
momenta of the parent system can appear at the nanotube zone center,
even though the primitive-cell $\Gamma$ point of the parent 2D
altermagnet is spin degenerate by symmetry. This distinction between
dimensional folding and curvature-induced modifications will become
particularly useful below when the nanotube bands are compared with
those of corresponding rectangular 2D supercells.

The continuous evolution between the limiting achiral orientations is
shown in Fig.~\ref{fig1}(f), where the spin-resolved energies at the
nanotube zone-boundary point $Z$ are plotted as a function of the chiral
angle $\theta$. For the $k_\perp=0$ subband, the spin splitting evolves continuously
from one antinodal orientation to the other, vanishes at
$\theta=45^\circ$, and reverses sign across the nodal direction,
consistent with Eq.~(\ref{eq:cos2theta}). The chiral angle therefore converts the
momentum-space anisotropy of the parent $d$-wave state into a
geometrically tunable property of the 1D electronic structure.

\subsection*{First-principles realization in V$_2$O nanotubes}

\begin{figure*}[ht!]
\centering
\includegraphics[width=0.63\textwidth]{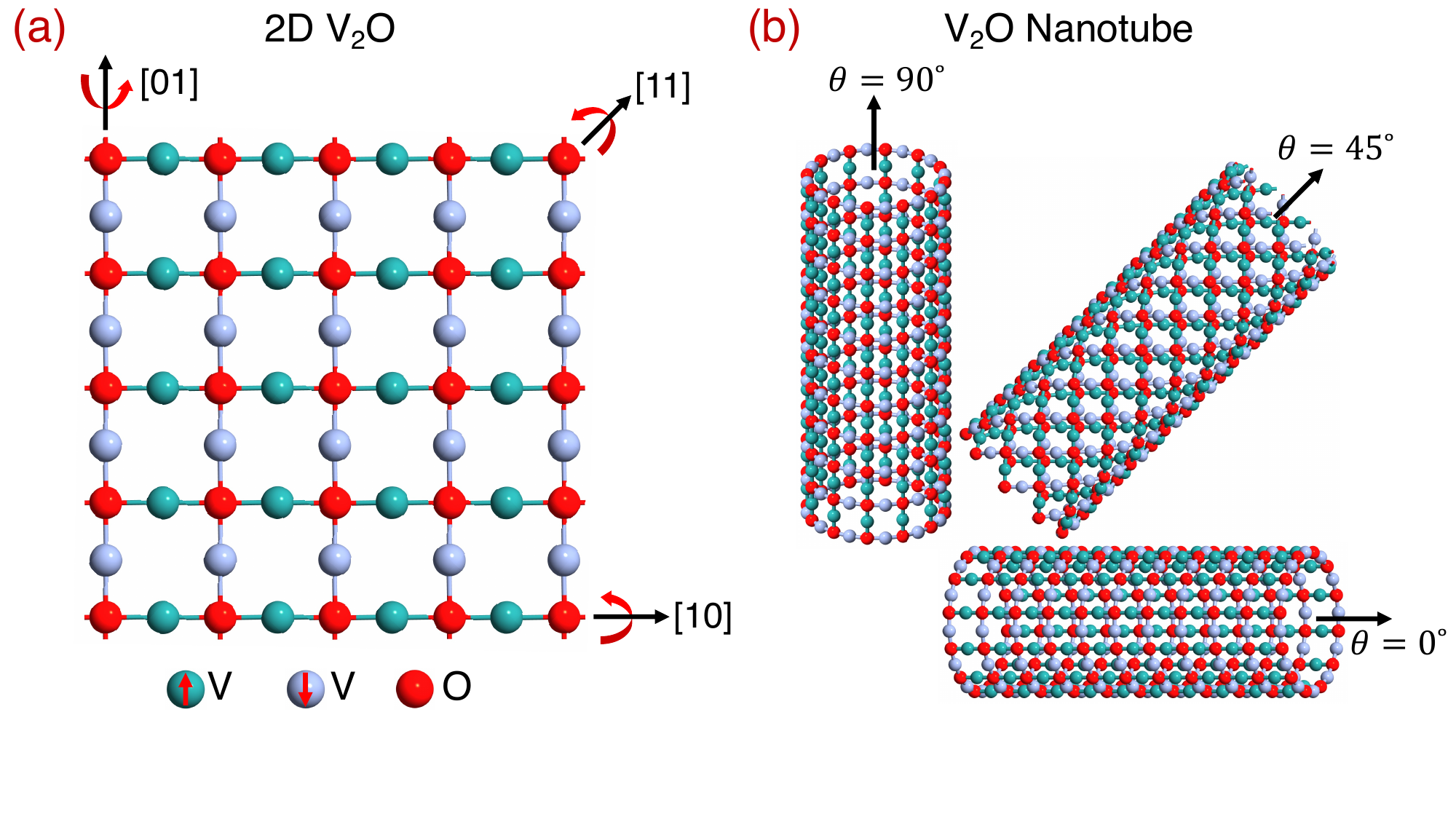}
\vspace{-0.9 cm}
\caption{
\textbf{Atomistic realization of nanotubes rolled from a 2D $d$-wave altermagnet}.
(a) Checkerboard 2D V$_2$O with opposite-spin V sublattices and
high-symmetry crystallographic directions $[10]$, $[11]$, and $[01]$.
(b) Representative achiral V$_2$O nanotubes with chiral angles
$\theta=0^\circ$, $45^\circ$, and $90^\circ$, defined by the
orientation of the nanotube axis relative to an antinodal direction of
the parent layer. The $\theta=0^\circ$ and $90^\circ$ geometries
correspond to antinodal orientations of the parent $d$-wave spin
splitting, whereas $\theta=45^\circ$ corresponds to the nodal
orientation.
}
\label{fig2}
\end{figure*}

We next examine how the dimensional-projection mechanism identified in
the tight-binding model is realized in an atomistic system. As a minimal
first-principles platform, we consider checkerboard V$_2$O, whose
square-lattice magnetic structure exhibits the $d_{x^2-y^2}$ spin
splitting required for the nodal--antinodal projection mechanism.
Figure~\ref{fig2}(a) shows the parent 2D V$_2$O layer together with the
high-symmetry crystallographic directions $[10]$, $[11]$, and $[01]$.
The $[10]$ and $[01]$ directions are antinodal, whereas the diagonal
$[11]$ direction is nodal.

Using the chiral-angle convention introduced above, the representative
nanotubes in Fig.~\ref{fig2}(b) correspond to $\theta=0^\circ$,
$45^\circ$, and $90^\circ$. These three limiting geometries are achiral
and provide atomistic realizations of the two orthogonal antinodal
orientations and the intervening nodal orientation identified in the
tight-binding model.

\begin{figure*}[t]
\centering
\includegraphics[width=0.99\textwidth]{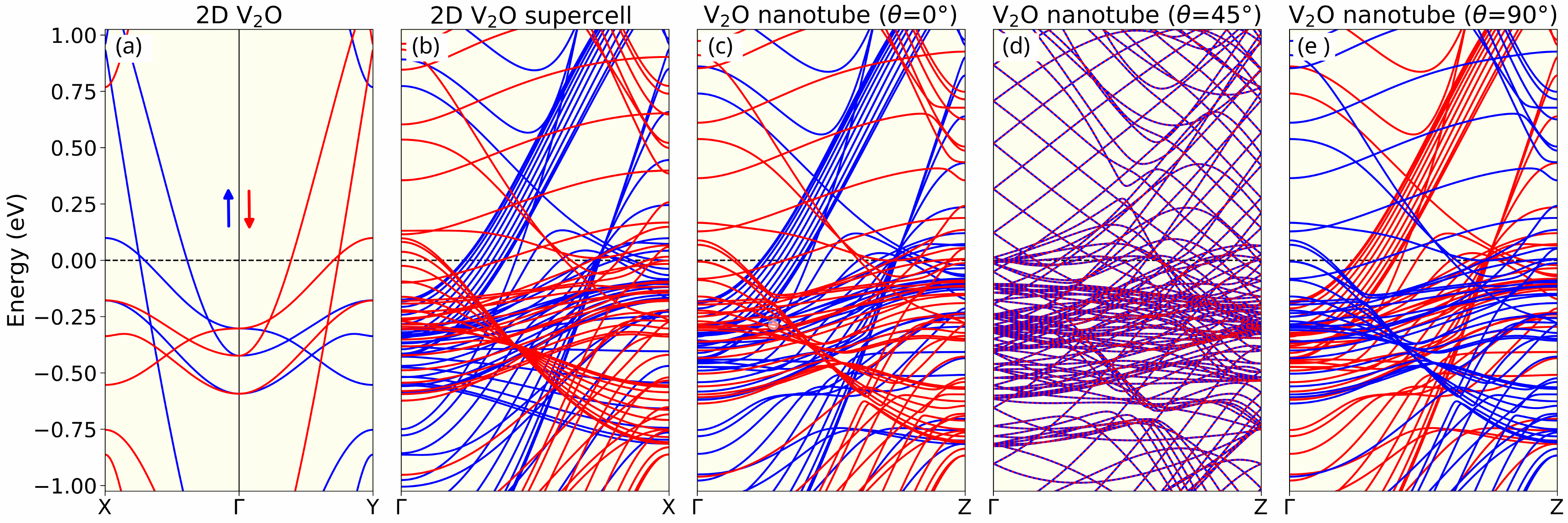}
\vspace{-0.4 cm}
\caption{
\textbf{First-principles dimensional projection of the V$_2$O electronic structure}.
(a) Spin-resolved band structure of primitive-cell 2D V$_2$O along
X--$\Gamma$--Y.
(b) Band structure of a $1\times18$ rectangular V$_2$O supercell along
$\Gamma$--X, showing the folded electronic structure prior to rolling.
(c)--(e) Spin-resolved band structures of the achiral V$_2$O nanotubes
with chiral angles $\theta=0^\circ$, $45^\circ$, and $90^\circ$,
respectively, calculated along the nanotube axis ($\Gamma$--Z). The two
antinodal nanotubes exhibit finite spin splitting with opposite signs,
whereas the nodal nanotube remains spin degenerate.
}
\label{fig3}
\end{figure*}

The corresponding first-principles electronic structure of the parent
V$_2$O monolayer is shown in Fig.~\ref{fig3}(a). Along the
X--$\Gamma$--Y path, the bands exhibit pronounced momentum-dependent
spin splitting away from the Brillouin-zone center, while the
$\Gamma$ point remains spin degenerate due to the symmetry of the
parent checkerboard lattice. The splitting reverses sign between the
orthogonal X--$\Gamma$ and $\Gamma$--Y directions, displaying the
characteristic $d_{x^2-y^2}$ structure anticipated by the tight-binding
model.

To separate the effects of dimensional folding from those introduced by
curvature, we first consider a $1\times18$ rectangular supercell of the
2D monolayer containing the same number of atoms as the corresponding
$\theta=0^\circ$ nanotube unit cell. Its band structure along
$\Gamma$--X, corresponding to the direction that becomes the nanotube
axis, is shown in Fig.~\ref{fig3}(b). The enlarged real-space
periodicity folds finite-momentum states of the primitive 2D system into
the reduced supercell Brillouin zone. Consequently, spin-split states
can appear at the supercell zone center even though the primitive-cell
$\Gamma$ point is spin degenerate. The rectangular supercell therefore
provides a curvature-free reference for identifying the electronic
changes associated with subsequent rolling.

The nanotube band structures for $\theta=0^\circ$, $45^\circ$, and
$90^\circ$ are shown in Figs.~\ref{fig3}(c)--\ref{fig3}(e),
respectively. The two antinodal nanotubes retain pronounced spin
splitting along the 1D $\Gamma$--Z direction, with opposite signs for
the two orthogonal orientations. In contrast, the $\theta=45^\circ$
nanotube remains spin degenerate, consistent with alignment of the
nanotube axis along a nodal direction of the parent $d$-wave state.
These first-principles band structures therefore reproduce the
nodal--antinodal dependence predicted by the tight-binding model.

The correspondence between the rectangular-supercell and nanotube band
structures further shows that the characteristic spin splitting is
already encoded in the dimensionally folded electronic structure rather
than being generated by curvature. Rolling modifies this electronic
structure through curvature and structural relaxation, but preserves
the underlying nodal--antinodal dependence. The relation between the
supercell and nanotube bands is further illustrated in Fig.~S1 of the
Supplementary Information.

The V$_2$O nanotubes also remain nearly magnetically compensated after
rolling. For a representative radius of $R=11.04$~\AA, the calculated
magnetic moments on the two V sublattices are $+2.63~\mu_B$ and
$-2.61~\mu_B$, corresponding to a residual moment imbalance
$\Delta M=|M_1+M_2|\simeq0.02~\mu_B$, while the O moments remain
negligible. The persistence of the chiral-angle-dependent spin splitting
in the presence of this small imbalance shows that the projected
electronic structure is robust against a slight departure from exact
magnetic compensation.

\subsection*{Chiral-angle-controlled spin transport}

Having established that the chiral angle controls the projected spin
splitting of the nanotube bands, we next examine how this geometrical
dependence is reflected in electronic transport. We calculate the
spin-resolved ballistic transmission along the nanotube axis for
V$_2$O nanotubes with chiral angles $\theta=0^\circ$,
$\simeq18^\circ$, $\simeq27^\circ$, and $45^\circ$. The
$\theta=0^\circ$ and $45^\circ$ nanotubes represent the achiral
antinodal and nodal limits, respectively, whereas the intermediate
$\theta\simeq18^\circ$ and $\simeq27^\circ$ nanotubes are chiral.
These structures therefore allow us to track how the spin polarization
of the conducting channels evolves from the antinodal to the nodal
orientation as the projected $d$-wave spin splitting is progressively
suppressed.

\begin{figure*}[t]
\centering
\includegraphics[width=0.999\textwidth]{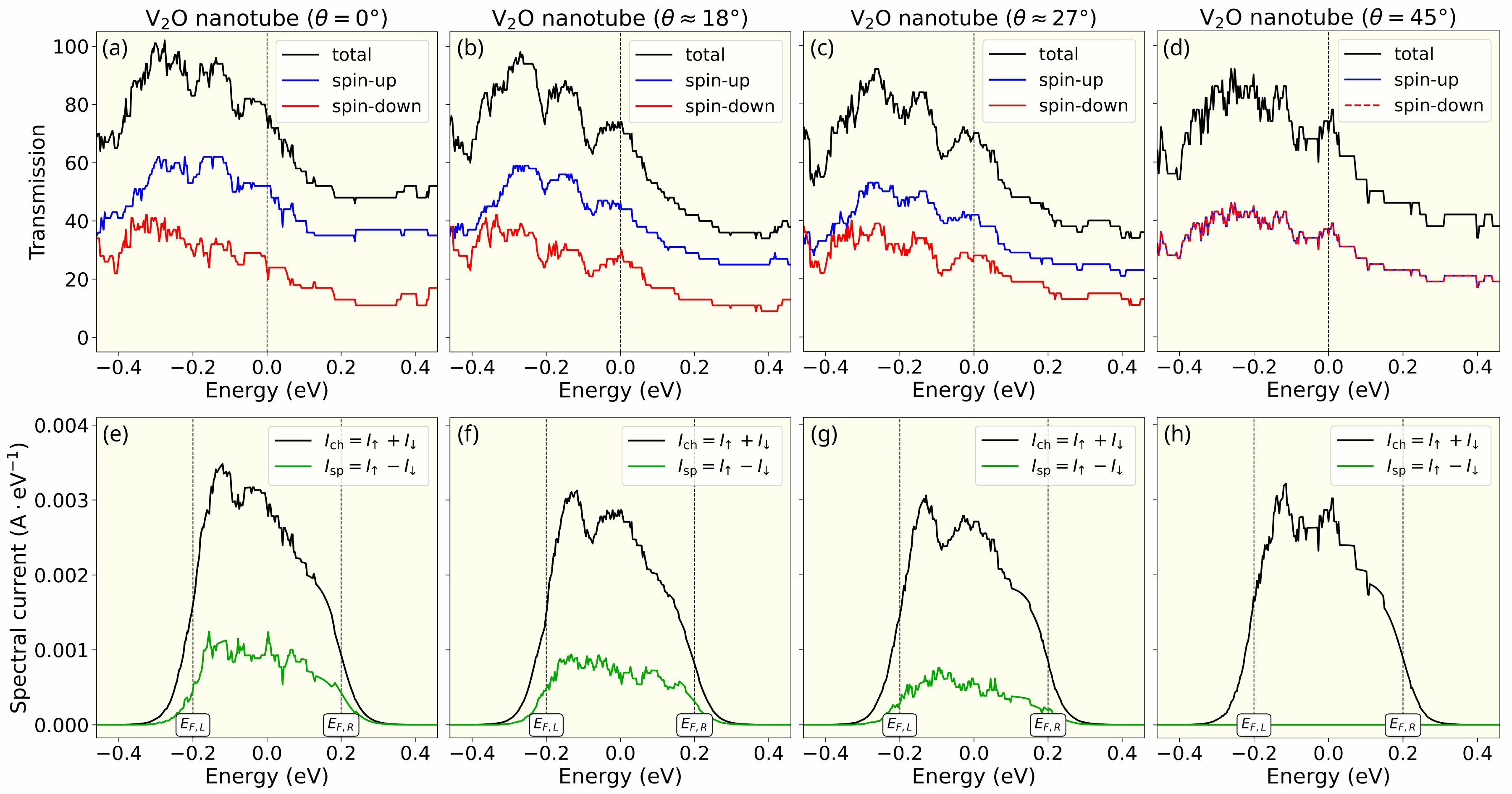}
\vspace{-0.3 cm}
\caption{
\textbf{Chiral-angle-controlled spin transport in V$_2$O nanotubes}.
(a)--(d) Zero-bias spin-resolved ballistic transmission spectra for
nanotubes with chiral angles $\theta=0^\circ$, $\simeq18^\circ$,
$\simeq27^\circ$, and $45^\circ$, respectively. The
$\theta=0^\circ$ and $45^\circ$ nanotubes are achiral, whereas the
intermediate-angle nanotubes are chiral. The total transmission and the
spin-up and spin-down contributions are shown as functions of energy,
with the equilibrium Fermi level set to zero. The transmission
polarization decreases on approaching the nodal direction and vanishes
for the $\theta=45^\circ$ nanotube.
(e)--(h) Energy-resolved spectral currents evaluated from the
corresponding zero-bias transmission spectra with electrochemical
potentials $E_{F,L}=-0.2$~eV and $E_{F,R}=+0.2$~eV relative to the
equilibrium Fermi level, defining a $0.4$~eV energy window. The charge
and spin spectral currents are defined as
$I_{\rm ch}=I_\uparrow+I_\downarrow$ and
$I_{\rm sp}=I_\uparrow-I_\downarrow$, respectively. The charge spectral
current remains of comparable magnitude for the different chiral
angles, whereas the spin spectral current decreases toward the nodal
orientation and vanishes at $\theta=45^\circ$.
}
\label{fig4}
\end{figure*}

Figures~\ref{fig4}(a)--\ref{fig4}(d) show the zero-bias total and
spin-resolved ballistic transmission spectra. For the antinodal
$\theta=0^\circ$ nanotube, the two spin channels exhibit a pronounced
imbalance around the Fermi level, with
$T_\uparrow(E_F)=52$ and $T_\downarrow(E_F)=23$, giving a total
transmission of $T(E_F)=75$. For $\theta\simeq18^\circ$ and
$\simeq27^\circ$, the corresponding spin-resolved values are $(46,28)$
and $(42,28)$, with total transmissions of $74$ and $70$, respectively.
At the nodal orientation, $\theta=45^\circ$, the two spin channels
become equivalent, $T_\uparrow(E_F)=T_\downarrow(E_F)=35$, while the
total transmission remains finite at $T(E_F)=70$. Thus, the
spin-channel imbalance is progressively suppressed on approaching the
nodal direction, whereas the overall number of conducting channels
remains comparable.

A convenient measure of the spin polarization of the transmission is
\begin{equation}
P_T(E)
=
\frac{T_\uparrow(E)-T_\downarrow(E)}
     {T_\uparrow(E)+T_\downarrow(E)}.
\label{eq:PT}
\end{equation}
At the Fermi level, $P_T$ decreases from approximately $38.7\%$ for
$\theta=0^\circ$ to $24.3\%$ and $20.0\%$ for
$\theta\simeq18^\circ$ and $\simeq27^\circ$, respectively, and
vanishes for $\theta=45^\circ$. The transport response therefore
preserves the nodal--antinodal trend identified in the electronic
structure: antinodal projections support a finite transmission
polarization, whereas the nodal projection restores equal contributions
from the two spin channels.

The transmission polarization should not be expected to follow the
simple $\cos(2\theta)$ dependence obtained for the $k_{\perp}=0$
subband. This scaling follows from the projected $d$-wave form factor
for this particular circumferential mode, whereas the transmission at
a given energy generally receives contributions from multiple
propagating subbands and depends on their dispersions and crossings
with that energy. Moreover, commensurate nanotubes with different
chiral angles generally differ slightly in radius and unit-cell
geometry, so the small variation of the total transmission from $70$
to $75$ cannot be attributed uniquely to the chiral angle. The robust
transport signature is instead the progressive reduction of the
spin-channel imbalance, from
$T_\uparrow-T_\downarrow=29$ at $\theta=0^\circ$ to zero at
$\theta=45^\circ$, while finite charge transmission is retained.

To examine the corresponding current response,
Figs.~\ref{fig4}(e)--\ref{fig4}(h) show the energy-resolved spectral
currents evaluated from the zero-bias transmission spectra. The left
and right electrochemical potentials were set to
$E_{F,L}=-0.2$~eV and $E_{F,R}=+0.2$~eV relative to the equilibrium
Fermi level, respectively, corresponding to a separation of
$0.4$~eV. The spin-resolved spectral current is
\begin{equation}
I_\sigma(E)
=
\frac{e}{h}
T_\sigma(E)
\left[
f_R(E)-f_L(E)
\right],
\label{eq:spectral_current}
\end{equation}
where $f_L$ and $f_R$ are the Fermi--Dirac distributions associated
with the left and right electrochemical potentials. Because
$T_\sigma(E)$ is taken from the zero-bias calculation and is not
recalculated self-consistently for the imposed electrochemical-potential
difference, these spectral currents represent a fixed zero-bias
transmission approximation rather than a self-consistent finite-bias
transport calculation. We define the charge and spin spectral currents
as
\begin{equation}
I_{\rm ch}(E)
=
I_\uparrow(E)+I_\downarrow(E),
\quad
I_{\rm sp}(E)
=
I_\uparrow(E)-I_\downarrow(E).
\label{eq:charge_spin_current}
\end{equation}
Consistent with the transmission spectra, the charge spectral current
remains of comparable magnitude for the different chiral angles,
whereas the spin spectral current decreases toward the nodal orientation
and vanishes at $\theta=45^\circ$, where the two spin-resolved
transmissions coincide.

The spectral-current results therefore connect the
chiral-angle-controlled electronic structure directly to spin-polarized
transport. Antinodal nanotubes support a finite spin spectral current,
whereas the nodal nanotube remains conducting but exhibits no spin
spectral current within the considered electrochemical-potential window.
Thus, the nodal--antinodal structure inherited from the parent 2D
$d$-wave altermagnet governs not only the projected 1D spin splitting
but also the spin polarization of the propagating channels.

\subsection*{Robustness across material realizations}

The V$_2$O nanotubes considered above provide a minimal realization of
the dimensional-projection mechanism, but an important question is
whether the same behavior persists in chemically and structurally more
complex systems. Rolling generally makes the inner and outer sides of a
nanotube inequivalent and can modify local bonding, hybridization, and
magnetic moments. Such effects are expected to become particularly
important in low-symmetry and Janus materials, where structural
asymmetry is already present in the parent monolayer. We therefore test
the robustness of the chiral-angle-controlled spin splitting in
representative symmetric and Janus nanotubes with different degrees of
curvature-induced magnetic imbalance.

The broader materials landscape provides several possible parent
systems for such realizations. Layered V--O--chalcogen compounds,
including KV$_2$Se$_2$O and RbV$_2$Te$_2$O, provide experimentally
realized systems with closely related magnetic and structural motifs
~\cite{Jiang2025_NatPhys_KV2Se2O,RbV2Te2O,AltermagnetV2XO}, while
recent computational searches have identified a wider range of
predicted 2D altermagnetic candidates
~\cite{xu2026chemical,gonzalez2026coexistence}. Representative parent
monolayers are listed in Table~S1 of the Supplementary Information.
These materials should be regarded as a candidate space rather than as
validated nanotube systems, since the stability and electronic structure
of each rolled geometry must be established explicitly.

To test the mechanism beyond V$_2$O, we performed first-principles
calculations for V$_2$OSe$_2$, V$_2$OSeTe, and Fe$_2$SSe.
V$_2$OSe$_2$ provides a symmetric reference, V$_2$OSeTe introduces
intrinsic Janus asymmetry, and Fe$_2$SSe represents a more strongly
perturbed Janus case in which rolling produces a substantially larger
magnetic moment imbalance. For each material, three representative nanotube
radii were considered to assess the effects of curvature on the magnetic
and electronic structure.

The calculated magnetic moments are summarized in Table~S2 of the
Supplementary Information. For V$_2$OSe$_2$, the residual moment
imbalance $\Delta M$ decreases from $0.15~\mu_B$ at
$R\simeq5$~\AA{} to $0.04~\mu_B$ at $R\simeq10$~\AA{} and
$0.02~\mu_B$ at $R\simeq15$~\AA{}. The Janus V$_2$OSeTe nanotubes
exhibit larger imbalances of $0.35~\mu_B$, $0.15~\mu_B$, and
$0.10~\mu_B$ over the same approximate radius range. An even stronger
imbalance is found for Fe$_2$SSe, where $\Delta M$ reaches
$0.69~\mu_B$ for the smallest nanotube and decreases to $0.42~\mu_B$
and $0.26~\mu_B$ at radii of approximately $10$ and $15$~\AA{},
respectively. The systematic decrease of $\Delta M$ with increasing
radius is consistent with the weakening of curvature-induced
inequivalence between the inner and outer sides of the nanotube.

Despite these substantial differences in magnetic imbalance, the
characteristic nodal--antinodal dependence of the electronic structure
remains preserved. Figure~S2 of the Supplementary Information compares
the parent monolayer, rectangular-supercell, and nanotube band
structures for representative V$_2$OSeTe and Fe$_2$SSe systems. For
V$_2$OSeTe, the $\theta=0^\circ$ nanotube retains pronounced spin
splitting and closely follows the folded electronic structure of the
corresponding rectangular supercell, whereas the $\theta=45^\circ$
nanotube remains spin degenerate along the nanotube axis. The same
qualitative distinction persists in Fe$_2$SSe despite the much larger
curvature-induced magnetic imbalance.

The Fe$_2$SSe bands also show that stronger curvature-induced structural
and magnetic modifications can lead to quantitative deviations from the
simple folded-supercell picture. These deviations, however, do not
eliminate the geometrical selection rule in the calculated cases:
antinodal orientations retain finite spin splitting, whereas the nodal
orientation remains spin degenerate. The dimensional-projection
mechanism is therefore not contingent on exact magnetic compensation or
on an electronically rigid rolling process. Rather, curvature-induced
changes act as perturbations to the nodal--antinodal projection imposed
by the momentum dependence of the parent $d$-wave spin splitting.

Overall, the V$_2$O, V$_2$OSe$_2$, V$_2$OSeTe, and Fe$_2$SSe
calculations show that the projected nodal--antinodal spin-splitting
pattern persists across structurally distinct nanotube realizations and
over a broad range of curvature-induced magnetic imbalance. These
results establish dimensional projection as a robust geometrical
mechanism for transferring the momentum-dependent spin splitting of 2D
$d$-wave altermagnets into controllable one-dimensional electronic
states.

\section*{Discussion}

An important consequence of dimensional reduction concerns the relation
between the symmetry of the parent 2D altermagnet and that of the
resulting nanotube. In the checkerboard 2D system, the combined
$C_4\mathcal{T}$ symmetry relates the two magnetic sublattices and
constrains the characteristic sign-changing $d$-wave spin-splitting
pattern between orthogonal momentum-space directions. Rolling the layer
into a nanotube generally breaks this global $C_4\mathcal{T}$ symmetry,
and the resulting 1D structure should therefore not, in general, be
regarded as an altermagnet in the same symmetry-classification sense as
its 2D parent. Nevertheless, the local magnetic environments and
exchange interactions responsible for the parent $d$-wave spin splitting
remain largely preserved upon rolling. Dimensional projection then
selects particular sectors of this inherited electronic structure,
allowing the nanotube bands to retain an altermagnet-like spin splitting
whose magnitude and sign depend on the chiral angle. The nanotube
therefore preserves a characteristic electronic fingerprint of the
parent altermagnet even though the corresponding global symmetry
relation of the 2D parent is no longer present in the rolled geometry.

This distinction also clarifies the role of curvature and magnetic
compensation. Rolling makes the inner and outer sides of the nanotube
inequivalent, modifying local bonding and magnetic moments and producing
a finite magnetic moment imbalance, particularly in the Janus systems.
These effects perturb, rather than generate, the chiral-angle dependence,
which is already present in the curvature-free dimensional-projection
picture. The persistence of the nodal--antinodal distinction across the
calculated nanotubes therefore shows that neither retention of the
parent's global $C_4\mathcal{T}$ symmetry nor exact magnetic compensation
of the final nanotube is required for the projected spin-splitting
pattern to survive.

Beyond the electronic structure, the transport calculations establish
that the inherited nodal--antinodal character also governs spin-polarized
conduction. A spin-split band structure does not by itself determine the
spin polarization of a current, because transport is governed by the
propagating states available at a particular energy. Our results show
that this nodal--antinodal information remains encoded in the propagating
channels, providing chiral-angle control over both the projected 1D spin
splitting and the spin polarization of the conducting states. This
geometrical control suggests potential spintronic applications, as
nanotubes with different chiral angles can exhibit finite or vanishing
spin polarization and could provide active elements in spin filters,
spin valves, or magnetic tunnel junctions. Particularly attractive are
compensated or nearly compensated nanotubes, where pronounced
spin-polarized transport can coexist with a vanishing or very small net
magnetic moment.

These prospective functionalities should, however, be viewed within the
scope of the present calculations. The behavior demonstrated here is
confirmed for several structurally distinct nanotube realizations,
whereas the broader set of 2D altermagnets identified in the
Supplementary Information should be regarded as a candidate space rather
than as validated nanotube systems. The stability, magnetic
configuration, and electronic structure of each rolled material must
therefore be established individually. 
Likewise, the present transport calculations describe zero-bias
ballistic transmission and spectral currents evaluated within a fixed
zero-bias transmission approximation, rather than a self-consistent
finite-bias transport calculation. Contacts, disorder, defects, finite-temperature 
broadening, and scattering may modify the resulting spin polarization in realistic
devices, and explicit device-level calculations will be required to
assess the proposed spintronic functionalities.

More broadly, our results illustrate how dimensional projection can
transfer characteristic electronic signatures of a higher-dimensional
magnetic state to a reduced-dimensional system even when the global
symmetry classification of the parent is not retained. In the present
case, the momentum-dependent $d$-wave spin splitting of a 2D altermagnet
becomes a geometrically tunable property of a 1D nanotube, with the
chiral angle controlling both the projected spin splitting and the spin
polarization of the conducting channels. Similar principles may apply to
nanoscrolls, curved ribbons, and other quasi-one-dimensional structures
formed from anisotropic magnetic layers, providing a route for translating symmetry-constrained
momentum-space electronic structure into geometrically controlled
low-dimensional electronic and transport properties.

\section*{Methods}

\subsection*{Tight-binding calculations}

The tight-binding calculations were performed using the minimal
two-sublattice model introduced in the main text. Nanotube states were
obtained by rotating the momentum coordinates according to the chiral
angle $\theta$ and imposing periodic boundary conditions along the
circumferential direction, resulting in quantized transverse momenta.
The model parameters and further numerical details are provided in the
Supplementary Information.

\subsection*{First-principles calculations}

First-principles calculations were performed within density-functional
theory using the generalized gradient approximation in the
Perdew--Burke--Ernzerhof (PBE) parametrization for the
exchange--correlation functional. All calculations were carried out
using the \textsc{QuantumATK} package with norm-conserving PseudoDojo
pseudopotentials and localized atomic-orbital basis sets
\cite{smidstrup2019an,perdew1996generalized,QuantumATKPseudoDojo}.
Collinear magnetic calculations were performed without spin--orbit
coupling. Atomic structures were relaxed until the residual forces were
below $0.01$~eV/\AA. A vacuum separation larger than $20$~\AA{} was
used between periodic images along the non-periodic directions.
Brillouin-zone integrations for the 2D monolayers employed
$20\times20\times1$ Monkhorst--Pack meshes, while the nanotubes were
sampled using at least 20 $k$ points along the one-dimensional
Brillouin zone. Nanotubes were constructed from commensurate supercells
of the corresponding parent monolayers with their axes aligned along the
desired crystallographic directions. Further computational and
structural details are provided in the Supplementary Information.

\subsection*{Spin-dependent transport}

Spin-resolved ballistic transmission spectra were calculated using
\textsc{QuantumATK} for the periodic bulk configurations of the
nanotubes, with the nanotube axis chosen as the transport direction.
No explicit device or scattering-region geometry was introduced. The
resulting transmission $T_\sigma(E)$ describes the number of
propagating channels of spin $\sigma$ available at a given energy along
the nanotube axis. All transmission spectra were calculated at zero
bias.

Energy-resolved spectral currents were subsequently obtained from the
zero-bias transmission spectra by setting the left and right
electrochemical potentials to $-0.2$ and $+0.2$~eV relative to the
equilibrium Fermi level, respectively, corresponding to an
electrochemical-potential difference of $0.4$~eV. Since the transmission
functions were not recalculated self-consistently at finite bias, the
spectral currents represent a fixed zero-bias transmission approximation
rather than a self-consistent finite-bias transport calculation.

\section*{Data availability}

The data supporting the findings of this study, including the numerical
data underlying the electronic-structure and transport results, are
available from the corresponding author upon reasonable request.

\section*{Code availability}

The first-principles and transport calculations were performed using
the commercially available QuantumATK software. The tight-binding model
and all parameters required to reproduce the calculations are provided
in the Supplementary Information. The custom codes used to implement
the tight-binding calculations are available from the corresponding
author upon reasonable request.

\section*{Acknowledgements}

This work was supported by the Deutsche Forschungsgemeinschaft (DFG)
through CRC/TRR 227, Project No. B12 (Project ID 328545488), and Grant
No. LO 1659/10-1, and by the European Innovation Council (EIC) Pathfinder
Open project, Grant No. 101129641. T.G.S. acknowledges computing time provided 
through the Gauss Centre for Supercomputing e.V. on the supercomputer JUWELS at
Forschungszentrum Jülich under project superorb.

\section*{Author contributions}

E.\c{S}. conceived the study and performed the first-principles and transport
calculations. T.G.S. and B.G. performed the tight-binding calculations and contributed 
to the theoretical analysis. I.M. and S.L. supervised the project. All authors
contributed substantially to the interpretation of the results. E.\c{S}. wrote
the manuscript with input from all authors.

\section*{Competing interests}

The authors declare no competing interests.

\bibliography{AM_Nanotubes}

\end{document}